\documentclass[%
 reprint,
 amsmath,amssymb,
 aps,
]{revtex4-2}

\usepackage{graphicx}
\usepackage{dcolumn}
\usepackage{bm}

\begin{document}

\preprint{APS/123-QED}

\title{Zeptometer displacement sensing using cavity opto-magneto-mechanics}

\author{Tatiana Iakovleva$^{1}$}
\email{tatiana.iakovleva@oist.jp}
\author{Bijita Sarma$^{1,2}$}
\author{Jason Twamley$^{1}$}
\affiliation{$^{1}$Quantum Machines Unit, Okinawa Institute of Science and Technology Graduate University, Okinawa 904-0495, Japan}
\affiliation{$^{2}$Department of Physics, Friedrich-Alexander-Universit\"at Erlangen-N\"urnberg, Staudtstra{\ss}e 7, 91058 Erlangen, Germany}

\date{\today}

\begin{abstract}
Optomechanical systems have been proven to be very useful for precision sensing of a variety of forces and effects. In this work, we propose an opto-magno-mechanical setup for spatial displacement sensing where one mirror of the optical cavity is levitated in vacuum via diamagnetic forces in an inhomogenous magnetic field produced by two layers of permanent magnets. We show that the optomechanical system can sense small changes in separation between the magnet layers $d$, via the small resulting shift in the mechanical frequency of the levitated mirror.
 We use Quantum Fisher Information (QFI) to quantify 
 the displacement sensing precision and study the fundamental precision bound that can be reached in our setup. Nonlinear interactions, that are inherently present in the optomechanical setup, improves the precision and we show that in the case of a pure state of the optical cavity one can achieve an extremely small displacement sensing precision of $\Delta d\sim36\times10^{-21}\text{m}$. We further incorporate decoherence to study its effect  on the fundamental precision attainable and find a relatively simple measrement protocol that can nearly achieve this fundamental precision limit.  
\end{abstract}

\maketitle


\section{\label{sec:level1}INTRODUCTION}

The ability to conduct precise measurements has always been important for scientific development. Every experiment designed to confirm a specific theory relies on measuring or estimating a parameter. 
The precision of measurement by classical means before the emergence of experimentally realizable quantum techniques was bounded by the so-called Standard Quantum Limit (SQL), with the error scaling as $1/\sqrt N$, where $N$ represents the resources used in the system such as the number of photons in optical sensing, or the number of times a measurement is repeated in specific cases. With the development of quantum theory, the SQL limit was surpassed \cite{Caves1981}, and an ultimate precision bound was achieved - the Heisenberg limit (HL), where the error scales as $1/N$ \cite{Zou2018,Mitchell2004}. This has opened up new prospects for development of ultraprecise devices and sensors, which in turn led to the discovery of gravitational waves detected in the LIGO experiment \cite{Abadie2011,Abbott2017}.

In recent years optomechanical systems have been proven useful for sensing purposes due to the inherent nonlinear coupling between photons and mechanical modes. 
In particular, it has been shown theoretically that this nonlinearity can be used to achieve sensitivity in gravitational acceleration measurements many orders of magnitude higher than atomic interferometers \cite{Szigeti2020a,Feng2020,Qvarfort2018,Xiao2020a}, and can also be used effectively for magnetometry \cite{Forstner2012,Zhu2022}, precise force sensing \cite{Zhang2022b,Zhao2020,Mason2019,Lee2022}, sideband cooling 
\cite{Arcizet2006,Schliesser2008,Gigan2006,Rossi2017}, and displacement sensing \cite{Harris2013,Mason2019,Liu2020,Sainadh2020,Peano2015}. These works use different quantum resources to improve sensing, such as exploitation of quantum correlations, injection of squeezed states of light and implementation of nonlinear optomechanical resonators etc.  

Here we theoretically propose and analyze a scheme for ultrahigh precision displacement sensing with a levitated opto-magno-mechanical setup, where a mirror is diamagnetically levitated in an inhomogeneous magnetic field generated by opposing magnets separated by a distance $d$. 
Levitated optomechanical systems consist of nano- or micro-scale objects confined in an optical, magnetic or electrical trap (Paul trap) \cite{Millen2020,Gonzalez2021}, which allows one to have precise control over the motion of the object. 

We consider an optical cavity formed between the magnetically levitated mirror and a stationary mirror kept above it, for the sensing of the displacement between the two magnets. The precision bound given by the Quantum Fisher Information (QFI) for the optical cavity field shows that the distance between the two magnets can be estimated with an ultra-high precision of $\Delta d\sim 36\times10^{-21}$m. In particular, we demonstrate that the variance scales as $\sim1/N^{1.5}$ for pure states of the optical cavity, where $|\psi\rangle=|\alpha\rangle$ with $|\alpha|^2\sim 10^7$ that can be achieved when one drives a high-Q optical cavity ($Q\sim10^5$), with a laser power $P\sim 40\,\text{nW}$.


\section{Optomechanical model}

\begin{figure*}
\centering
	\includegraphics[width=0.9\linewidth]{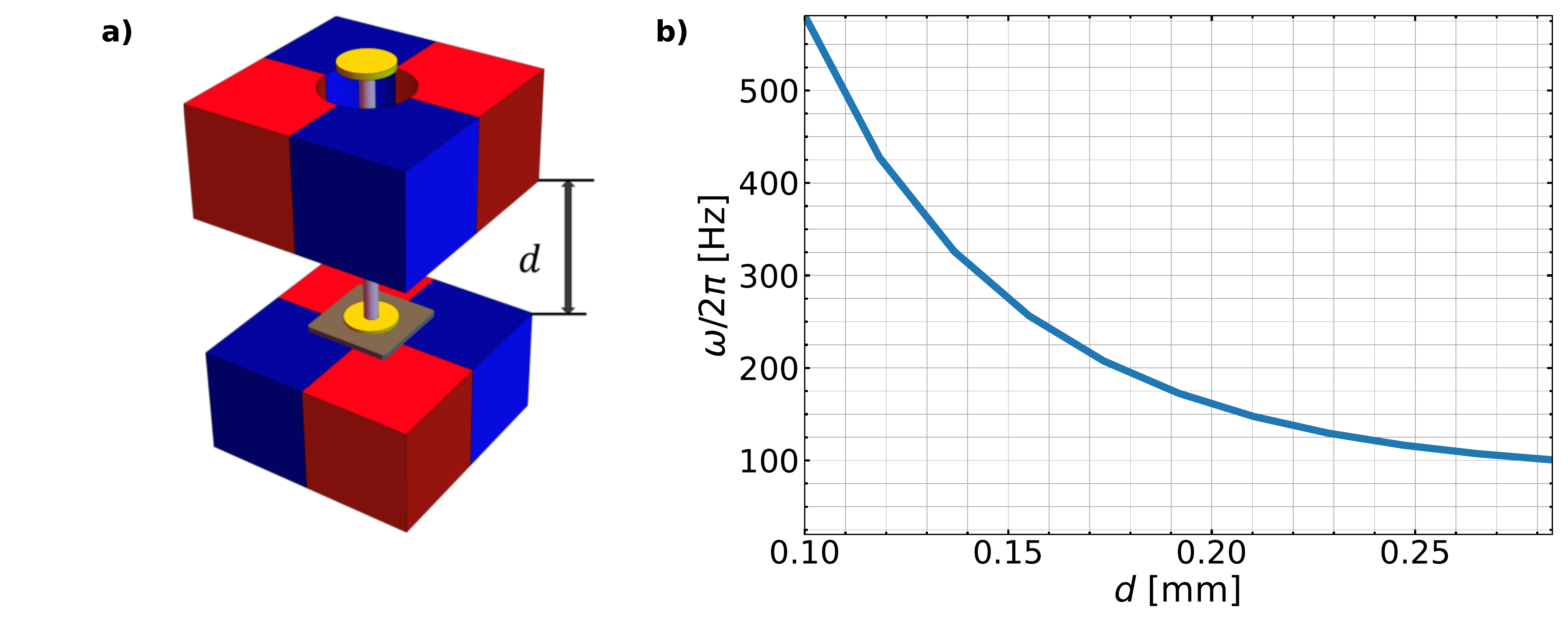}
\vspace*{-10pt}\caption{a) Proposed setup: a diamagnetic plate is levitated between two planar checkerboard magnetic arrays - the top one having a small clear central opening to admit an optical mode. The diamagnetic plate is harmonically trapped in all three dimensions. We are primarily interested in harmonic oscillations of the plate in the vertical direction. An optical cavity is formed by a mirror attached to the diamagnetic levitated plate and a fixed mirror above. We consider $\omega_c$  to be the optical cavity frequency, while $\omega(d)$ is the vertical mechanical oscillation frequency, which depends on the magnet-array separation distance $d$. As the separation $d$ changes between the magnet arrays the magnetic restoring force on the levitated diamagnetic plate alters  and thus the mechanical trap frequency $\omega$ depends on $d$. b) This plot shows how the vertical motional frequency of the levitated diamagnetic plate changes as a function of the separation $d$ between the magnet-arrays when we choose physical parameter values as in Table \ref{tab:param}. We note that there exists a critical value of $d$, denoted as $d_{\rm crit}$. For $d<d_{\rm crit}$, the levitated plate's equilibrium configuration is aligned with the magnet-array, while for $d>d_{\rm crit}$, the diamagnetic plate's equilibrium configuration is rotated by $\phi=\pi/4$, around the $z-$axis relative to the magnet-arrays. In b) we only show the domain when $d<d_{\rm crit}$. Mainly the vertical motional mode is of interest in this work, and it can be seen that there is a large change in the trapping frequency as the magnet-array separation distance is altered. 
}
\label{fig:scheme}
\end{figure*}

In this work, we consider a levitated opto-magno-mechanical setup as depicted in Fig.~\ref{fig:scheme}, where a mirror is diamagnetically levitated in an inhomogeneous magnetic field generated by two opposing  magnets separated by a distance $d$. Diamagnetic levitation can be achieved at room temperature by placing the mirror on diamagnetic graphite trapped between checkerboard magnet arrays \cite{Romagnoli2022}, or at low temperatures by placing the mirror on Type I or II superconductors whose diamagnetism is six orders of magnitude larger than that of graphite 
\cite{Johansson2013,Waldron1966}. This setup allows the mechanical frequency of the mirror oscillation to depend on the separation distance, $d$, between the magnet layers. Therefore, by measuring the oscillation frequency $\omega$ of the mirror, one can estimate the distance between the two magnet layers with high precision. We complement this setup with an optical cavity formed between the levitated mirror and a stationary mirror above it. 

The total Hamiltonian of the optomechanical system formed by the levitated and fixed mirrors and the cavity optical field is given by,

\begin{equation}\label{eq:1}
    H = \hbar\omega_ca^\dag a + \frac{p^2}{2m}+\frac{m\omega^2z^2}{2}-\hbar\frac{\omega_c}{L}a^\dag az - mgz,
\end{equation}

\noindent where $a^\dag, a$ are the creation and annihilation operators of the optical cavity, $g$ is the gravitational acceleration, $m$ is the mass of the mirror, $\omega_c$ is the cavity frequency and $\omega(d)$ is the mechanical center of mass frequency of the levitated mirror, which depends on the magnet separation distance $d$. The position and momentum operators are defined as follows: $z=\sqrt{\hbar/2m\omega}(b^\dag+b)$, $p=i\sqrt{m\omega\hbar/2}(b^\dag-b)$, where $b^\dag, b$ are the creation and annihilation operators of the motional mode of the levitated mirror. Applying the polaron transformation, $U_p=\text{exp}[(\chi a^\dag a-S)/\omega(b^\dag-b)]$ \cite{Mancini1997,Bose1997} and making use of the Baker-Campbell-Hausdorff formula \cite{Louisell1973} one can transform the Hamiltonian $H$ to the following form,

\begin{equation} \label{eq:2}
     H=\hbar\omega_ca^\dag a+\hbar\omega b^\dag b -\hbar (\chi a^\dag a + S)^2/\omega,
\end{equation}

\noindent where the single photon optomechanical coupling strength is given by $\chi=\frac{\omega_c}{L}\sqrt{\hbar/(2\omega m)}$, and $S=mg\sqrt{1/(2\omega\hbar m)}$ is the normalized constant containing the gravitational acceleration. After this transformation the interaction term implies a Kerr-type third-order nonlinear behavior $H_{\mathrm{int}}\sim\chi^2(a^\dag a)^2$, with the corresponding propagator: 
\begin{widetext}
\begin{equation}\label{eq:3}
    U(t)=e^{-i\omega_ct a^\dag a}e^{i(\chi a^\dag a+S)^2/\omega^2(t\omega-\sin(\omega t))}e^{(\chi a^\dag a+S)/\omega(b^\dag\eta-b\eta^*)}e^{-it\omega b^\dag b},
\end{equation}
\end{widetext}
\noindent where $\eta = 1-e^{-i\omega t}$. 

\section{Quantum estimation limits} 
To estimate  small changes in the separation $d$, of the two magnet layers which trap the levitated mirror, i.e.~$d\rightarrow d+\Delta d$, where $\Delta d \ll d$, one can apply the Quantum Cramer-Rao bound(QCRB) that gives the lowest boundary on the variance, 

\begin{equation}\label{eq:4}
    Var(\phi)\geq 1/(MF_Q),
\end{equation}
\noindent
where $F_Q$ is the QFI, and $M$ is the number of independent repetitions. From this expression one can conclude that an increase in the QFI will result in a decrease in the variance, and as a consequence more precise parameter estimation with smaller error. In the most general case, the QFI can be expressed via the Symmetric Logarithmic Derivative (SLD), $L_\phi$ as:
\begin{equation}\label{eq:qfi-sld}
    F_Q=\text{Tr}[\partial_\phi\rho_\phi L_\phi],
\end{equation}
\noindent where $\phi$ is the parameter we wish to estimate. 

Calculation of the QFI is rather a challenging task without using any features of the system such as an eigenbasis decomposition of the density matrix, or if the system is not in a pure state. In our case, since at certain times the mechanics and optics decouple, we can consider only the optical part of the system, and at these decoupling times (initially without considering any loss), this optical part will be in a pure state $\Psi(0)=|\alpha\rangle$.
 The QFI for a pure state \cite{Paris2009}, $\rho=|\Psi\rangle\langle\Psi|$, is given by,
\begin{equation} \label{eq:5}
    F_Q = 4[\langle\partial_\phi\Psi|\partial_\phi\Psi\rangle-|\langle\partial_\phi\Psi|\Psi\rangle|^2],
\end{equation}
\noindent
and, for a mixed state one can use eigenbasis decomposition of the density matrix, $\rho=\sum_i p_i|\psi_i\rangle\langle\psi_i|$, so that the QFI is given by,
\begin{equation} \label{eq:6}
    F_Q = 2\sum_{i,j}\frac{|\langle\psi_i|\partial_\phi\rho|\psi_j\rangle|^2}{p_i+p_j},
\end{equation}
where $|\psi_{i/j}\rangle$ is the eigenbasis of the density matrix $\rho$, and $p_{i/j}$'s are the eigenvalues, such that $p_i+p_j \neq 0$.
The parameter values used throughout the work are presented in Table \ref{tab:param}. 

\section{Analysis}

We consider the initial combined optomechanical state of the system formed by the intracavity optical mode and the  motional phonon mode of the movable levitated mirror as a separable product state of the form $\rho(0)=|\alpha\rangle\langle\alpha|\otimes\rho_{\text{th}}$, where $|\alpha\rangle$ is an optical coherent state and $\rho_{\mathrm{th}}=(1-e^{-\hbar\beta\omega})\sum_ne^{-\hbar\beta\omega n}|n\rangle\langle n|$ is the thermal state of the mechanical oscillator at temperature $T$, where $\beta=1/T$. 
After applying the evolution operator $U$ for a time $t$, we see that the system evolves into the entangled state,

\begin{widetext}
    \begin{equation}\label{eq:7}
        \rho(t)=e^{-|\alpha|^2}\sum_{l,m}\frac{\alpha^l\alpha^{*\,m}}{\sqrt{l!m!}}e^{i(\omega t-\sin(\omega t))(\chi^2(l^2-m^2)+2S\chi(l-m))/\omega^2}|l\rangle\langle m|\otimes D(\eta,l)\rho_{\text{th}}D^\dag(\eta,m),
    \end{equation}
\end{widetext}

\noindent where $D(\eta,l)=e^{(\chi l+S)/\omega(b^\dag\eta-b\eta^*)}$ is the displacement operator which depends on the number of photons $\hat{l}=a^\dag a$, with $\hat{l}|l\rangle=l|l\rangle$. 

However, we see that at $t_n=2n\pi/\omega$, where $n$ is an integer, the mechanical oscillator and optical cavity become decoupled:
\begin{multline}
    \label{eq:8}
    \rho(t_n)=e^{-|\alpha|^2}\sum_{l,m}\frac{\alpha^l\alpha^{*\,m}}{\sqrt{l!m!}}|l\rangle\langle m|\otimes \rho_{\text{th}}\times \\
    \times e^{2\pi i(\chi^2(l^2-m^2)+2S\chi(l-m))/\omega^2}
    .
\end{multline}

\begin{figure}
\centering
	\includegraphics[width=\linewidth]{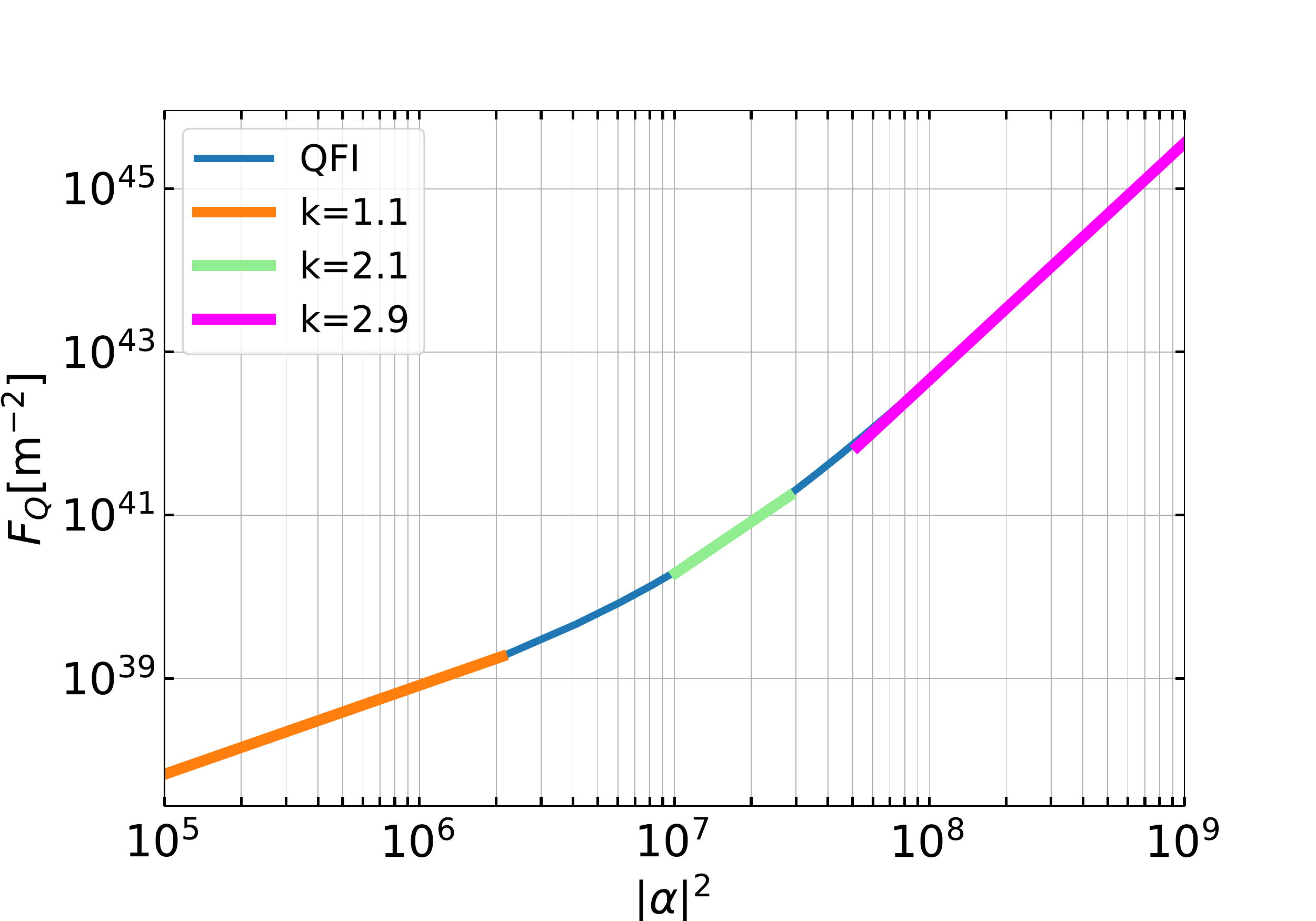}
\vspace*{-10pt}\caption{ This plot shows the QFI calculated at a fixed time point $t=2\pi/\omega$, when optical and mechanical components are decoupled and one can consider the optical cavity to be in a pure state. QFI in a log scale for the pure state of the cavity is calculated using Eq.~(\ref{eq:9}). The parameters used for calculation can be found in Table \ref{tab:param}. QFI scales as $F_Q\sim N^k$, where $N = |\alpha|^2$ is the number of photons in the cavity, and $k$ is the exponent determined from a numerical fit. The orange line shows approximate SQL scaling of the error $\Delta d\sim1/\sqrt{N}$ in the region up to $|\alpha|^2\sim10^7$, while the magenta line indicates scaling of the error $\Delta d\sim1/N^{1.5}$ and green line shows the transition region where the error scales as $\Delta d\sim1/N$.}
\label{fig:qfi-pure-states}
\end{figure}

At these decoupling times, the state of the optical cavity mode is pure, ignoring any optical and mechanical loss. Hence this makes it feasible to use the optical mode for parameter estimation using the QFI formula for pure states. We now seek to  obtain the limit for sensing small changes in $d$ as:

\begin{multline}
     \label{eq:9}
        F_Q(d) =4\Big(\frac{6\pi n }{\omega^3}\Big)^2|\alpha|^2\chi^2(6\chi^2|\alpha|^2+4\chi^2|\alpha|^4+\\
        +(\chi+2S)^2+8\chi S|\alpha|^2)\Big(\frac{d\omega}{d d}\Big)^2,
\end{multline}

\noindent where we use the QFI to return the precision to sense small changes in $\omega$, and using $d\omega/dd$, to convert this to small changes in $d$. For realistic experimental parameter values (Table \ref{tab:param}) we find the absolute value of the QFI as $F_{\rm Q}\sim7.6\times10^{38}\text{m}^{-2}$, with a sensitivity in estimating the distance $d$, to be $\Delta d\sim 36\times10^{-21} \rm m$. It is also interesting to look at the scaling with the number of photons, $N\sim|\alpha|^2$. In Fig.~\ref{fig:qfi-pure-states} it is shown that in the region where $|\alpha|^2 \geq10^7$ the scaling is $\Delta d\sim1/N^{1.5}$. 


Next we consider the times $t\neq 2\pi/\omega$, when the two subsystems remain coupled. In order to calculate the QFI according to Eq.~\ref{eq:6} we numerically simulate the system dynamics \cite{Johansson2013}, and compare the results with an analytical solution. Due to high computational overhead, these numerics can only treat cases of weak cavity fields where $|\alpha|^2\sim 1$. To obtain an analytical expression we choose to re-express the initial thermal state of the mechanics in a new form. The coherent basis representation or $P$-function representation of a thermal state is given by \cite{mandel_wolf_1995},
\begin{equation}\label{eq:10}
    \rho_{\text{th}}=\frac{1}{n_\beta\pi}\int d^2\gamma e^{-|\gamma|^2/n_\beta}|\gamma \rangle\langle\gamma|,
\end{equation}

\noindent where $n_\beta=1/(e^{\hbar\omega\beta}-1)$. Using the propagator from Eq.~\ref{eq:3} and the initial state $\rho_0=|\alpha\rangle\langle\alpha|\otimes\rho_{\text{th}}$, and using the above form for the thermal motional state, we obtain,

\begin{widetext}
    \begin{multline}\label{eq:11}
        \rho(t)=U(t)\rho_0U^\dag(t)=e^{-i\omega_c a^\dag at}e^{(\frac{\chi}{\omega} a^\dag a+\frac{S}{\omega})(b^\dag\eta-b\eta^*)}e^{-ib^\dag b\omega t}e^{i(\frac{\chi}{\omega}a^\dag a+\frac{S}{\omega})^2(t\omega-\sin(\omega t))}\\
        |\alpha\rangle\langle\alpha|\otimes\rho_{\text{th}}
        e^{i\omega_c a^\dag at}e^{-(\frac{\chi}{\omega} a^\dag a+\frac{S}{\omega})(b^\dag\eta-b\eta^*)}e^{ib^\dag b\omega t}e^{-i(\frac{\chi}{\omega}a^\dag a+\frac{S}{\omega})^2(t\omega-\sin(\omega t))}.
    \end{multline}

\noindent Tracing out the mechanical mode we get the reduced density matrix for the optical cavity as,
\begin{equation}
    \label{eq:12}
     \rho_c(t)=e^{-|\alpha|^2}\sum_{lm}\frac{\alpha^l\alpha^{*m}}{\sqrt{l!m!}}|l\rangle\langle m|e^{i(\chi^2/\omega^2(l^2-m^2)+2\chi s/\omega^2(l-m))(t\omega-\sin(\omega t))} e^{-\chi^2(l-m)^2/\omega^2(1-\cos(\omega t))(1+n_\beta/2)}.
\end{equation}
\end{widetext}
\noindent In Eq.~\ref{eq:12} the sum runs from 0 to infinity, and in order to numerically simulate this state we have to truncate it to a certain reasonable number of Fock states. This leads to loss of some information about the system. Another numerical obstacle lies in the parameters range that can be used for simulation. In order to simulate the system with $\alpha=1$, a Hilbert space dimension no less than $dim(\mathcal{H})=50$ is required, which leads to manipulation of $(2500\times2500)$-sized matrices. This restricts us to use only small values of $\alpha$ and other parameters for the numerical simulation. 

\begin{figure}
\centering
	\includegraphics[width=\linewidth]{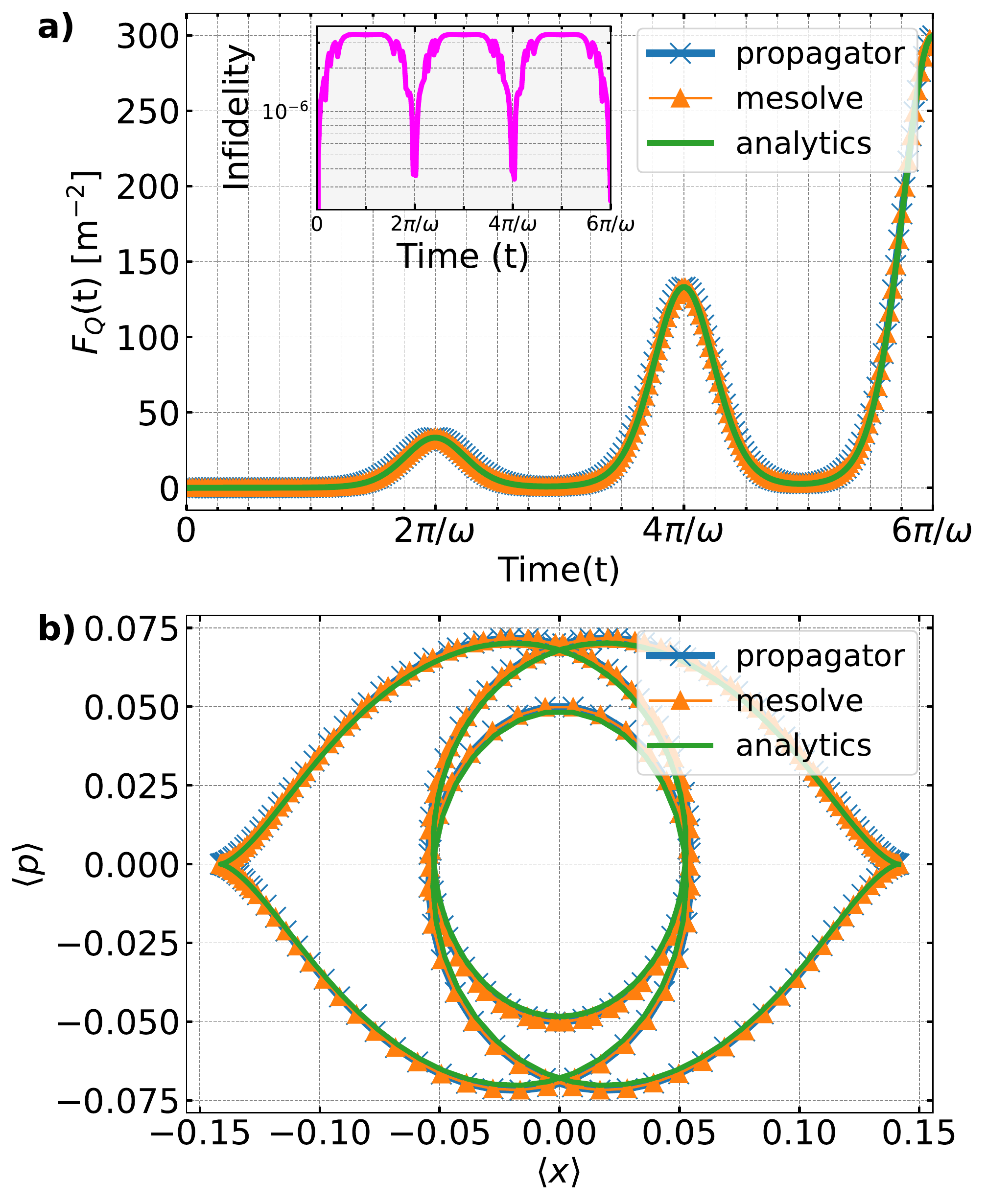}
\caption{a) Quantum Fisher Information evolution in time, with the comparison of analytical and numerical results. We use \textit{mesolve} method of the QuTiP package for numerical simulation. QFI achieves peak values at times $t=2\pi n/\omega$ when optical and mechanical parts decouple. The inset shows the infidelity between numerical and analytical states. b) Field quadratures $\langle x\rangle = (1/\sqrt{2})\langle a^\dag+a\rangle$ and $\langle p\rangle=(i/\sqrt{2})\langle a^\dag-a\rangle$ for the cavity optical mode traced out complex trajectories in the phase space, with $|\alpha|^2=0.01, n_\beta=0.1$, $t\in [0,6\pi/\omega]$.}
\label{fig:num-analyt}
\end{figure}

We compare the density matrices obtained numerically (using \textit{mesolve} method within the python package QuTiP \cite{Johansson2013}, and evolving state under the propagator in Eq.~\ref{eq:3}) and analytically, and conclude that both methods give the same result up to small numerical errors (Fig.~\ref{fig:num-analyt}(a)). We calculate the field quadratures $\langle x\rangle$ and $\langle p\rangle$ to monitor that no decoherence of the phase space trajectory is present due to the numerical simulation errors (Fig.~\ref{fig:num-analyt}(b)). Moreover, we also track the infidelity between the numerical and analytical density matrices, given by \cite{nielsen_chuang_2010}:
\begin{equation}
    \text{Infidelity}(\rho,\sigma) = 1 - \text{Tr}\sqrt{\sqrt{\rho}\sigma\sqrt{\rho}}.
\end{equation}
We find that the infidelity oscillates in time (shown in Fig.~\ref{fig:num-analyt}(a) inset). Between the decoupling times fidelity reaches minimum values, persisting throughout the time evolution. The minimum fidelity value is found to be $min\{\text{Fidelity}\} = 0.999996$, which we believe is sufficient to conclude that the numerical simulation gives the correct result.

\subsection{QFI in presence of loss}

Next, we consider a more realistic scenario that includes cavity loss, and we study the QFI behavior under decoherence which we model as repeated interactions with the external environment - also known as the {\it collision decoherence model} \cite{Ciccarello2022}. For this, we discretize the evolution by dividing it into smaller time steps $U(t)=\underbrace{U(\Delta t)U(\Delta t)...}_{N}$, and assume that after each time interval $\Delta t$ the system interacts and becomes entangled with an environment mode $E$ which is then traced out. This tracing out results in a loss of information in the system and simulates the action of the environment onto the system. The full quantum channel, that maps the initial state at time step $n-1$, $\rho_{n-1}$, to the next time step state $\rho_n$, and which simulates photons leaking from the cavity over the duration $\Delta t$, is given by,

\begin{multline}
    \Phi:\rho_{n-1}\rightarrow \rho_{n},\\
    \;\rho_n=\text{Tr}_{\text{E}}[U_{\text{BS}}(U(\Delta t)\rho_{n-1}U^\dag(\Delta t)\otimes |0\rangle\langle0|)U_{\text{BS}}^\dag]\;\;.
\end{multline}
\begin{figure}
\centering
	\includegraphics[width=1.0\linewidth]{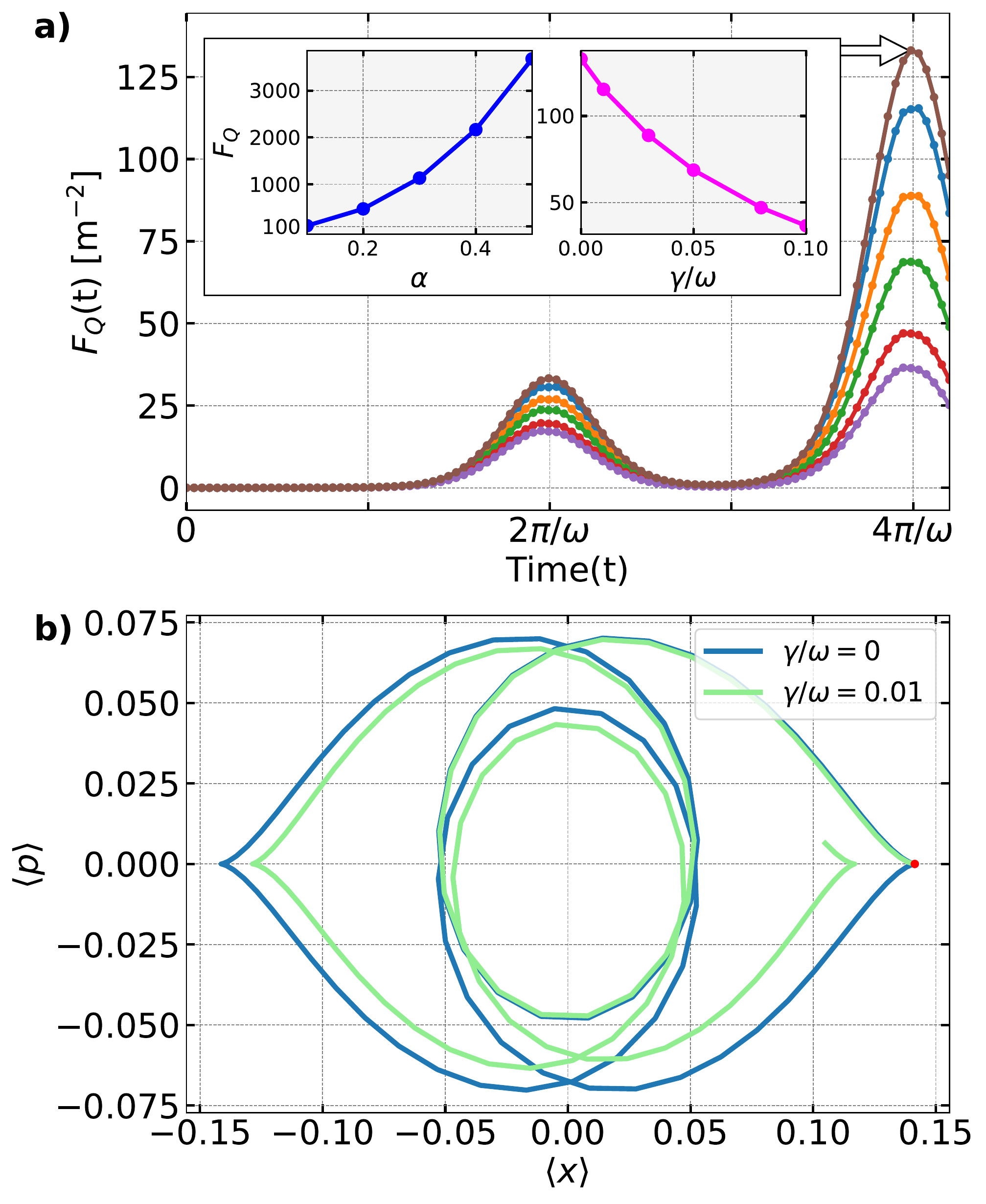}
\caption{a) Quantum Fisher Information evolution in time subjected to decoherence with decoherence rate $\gamma/\omega=\{0,0.01,0.03,0.05,0.08,0.1\}$ : (brown, blue, orange, green, red, purple) . QFI achieves peak values at times $t=2\pi n/\omega$ when optical and mechanical parts decouple. In the inset, the left plot shows the QFI at time $t=4\pi/\omega$(second peak) depending on the number of photons $\alpha=\{0.1,0.2,0.3,0.4,0.5\}$ for the same decoherence rate $\gamma/\omega=0.01$, whereas the right-side plot shows the value of the QFI at the same time point with respect to the decoherence rate $\gamma/\omega$ for $\alpha=0.1$. b) Field quadratures $\langle x\rangle$ and $\langle p\rangle$ for the cavity optical mode. Red dot indicates the starting point $t=0$. Other parameters are, $|\alpha|^2=0.01, n_\beta=0.1$, $t\in \{0,4.2\pi/\omega\}$, and the beam-splitter ratio is given by $\phi_\tau=0.036$.}
\label{fig:qfi-dec}
\end{figure}

This map starts with the density matrix $\rho_{n-1}$, of the total optomechanical system at time step $t_{n-1}$ which is initially in a product state with the ancilla environment mode, the latter prepared in the vacuum state $|0\rangle\langle 0|$. The map then evolves the optomechanical system one step forward in time via the optomechanical propagator $U(\Delta t)$. Then the map entangles the optomechanical system with the environmental mode via the application of a unitary beam splitter operation $U_{\text{BS}}=e^{-i\phi_\tau(a^\dag c+ ac^\dag)}$, where $a, c$ are the annihilation operators of the cavity optical mode and the environment mode respectively, and $\phi_\tau$ is the beam-splitter transmission rate. This beam splitter operation very slightly entangles the environment mode to the optical cavity mode. 
By taking a limit one can make a correspondence between the collision model and a Lindblad master equation description of the decoherence \cite{Ciccarello2022}. The beam-splitter parameters are related to the master equation decoherence rate through the relation $\phi_\tau=\sqrt{\gamma\Delta t}$, where $\Delta t$ is the discretization timestep, and $\gamma$ is the decoherence rate in Lindblad Master equation $\dot{\rho}=\gamma(a\rho a^\dag-1/2\{a^\dag a,\rho\})$ \cite{Ciccarello2022}. After this entangling operation,  we trace out the ancillary environment mode. The map $\Phi$, is iterated repeatedly in order to obtain the mixed state of the optomechanical system at a later time $t_N$, as $\rho_N=\Phi^N[\rho_0]$.

Fig.~\ref{fig:qfi-dec}(a) compares the QFI values without decoherence and in presence of decoherence with various values of $\gamma/\omega$. However, this simulation can only deal with small values for the energy in the optical cavity $\sim|\alpha|^2\ll 1$ as compared with Fig. \ref{fig:qfi-pure-states}, as we make use of the full system density matrix in the collision decoherence model. Fig.~\ref{fig:qfi-dec}(a) shows that at the decoupling times, the peak value of the QFI decreases with an increase in the value of $\gamma/\omega$. For example, the peak value of the QFI decreases compared to the zero decoherence case by a scale factor of $1.2$ for $\gamma/\omega=0.01$ at the second decoupling time $t=4\pi/\omega$. 
This difference grows in time and therefore the information that can be extracted from the system decreases with increasing loss. However, from the inset of Fig. \ref{fig:qfi-dec}(a), one can see that for a fixed loss rate $\gamma/\omega$, the QFI at the first peak increases with an increase in $\alpha$. If we double $\alpha$ then the second peak QFI more than doubles. This suggests that introducing small loss rates with higher photon number would not significantly reduce the QFI's observed in Fig.\ref{fig:qfi-pure-states}, and hence the precision scalings observed may still hold.

\subsection{Saturation of QFI}
It was shown that the optimal measurement saturating the QCRB (Eq.~\ref{eq:4}) is the projective measurement on the eigenbasis of the symmetric logarithmic derivative operator defined by $\partial\rho(d)/\partial d = 1/2(L\rho(d)+\rho(d)L)$ \cite{Caves1994}. However, the experimental implementation of such measurement in the case of mixed state is unclear. 

\begin{figure}[h!]
\centering
	\includegraphics[width=1.0\linewidth]{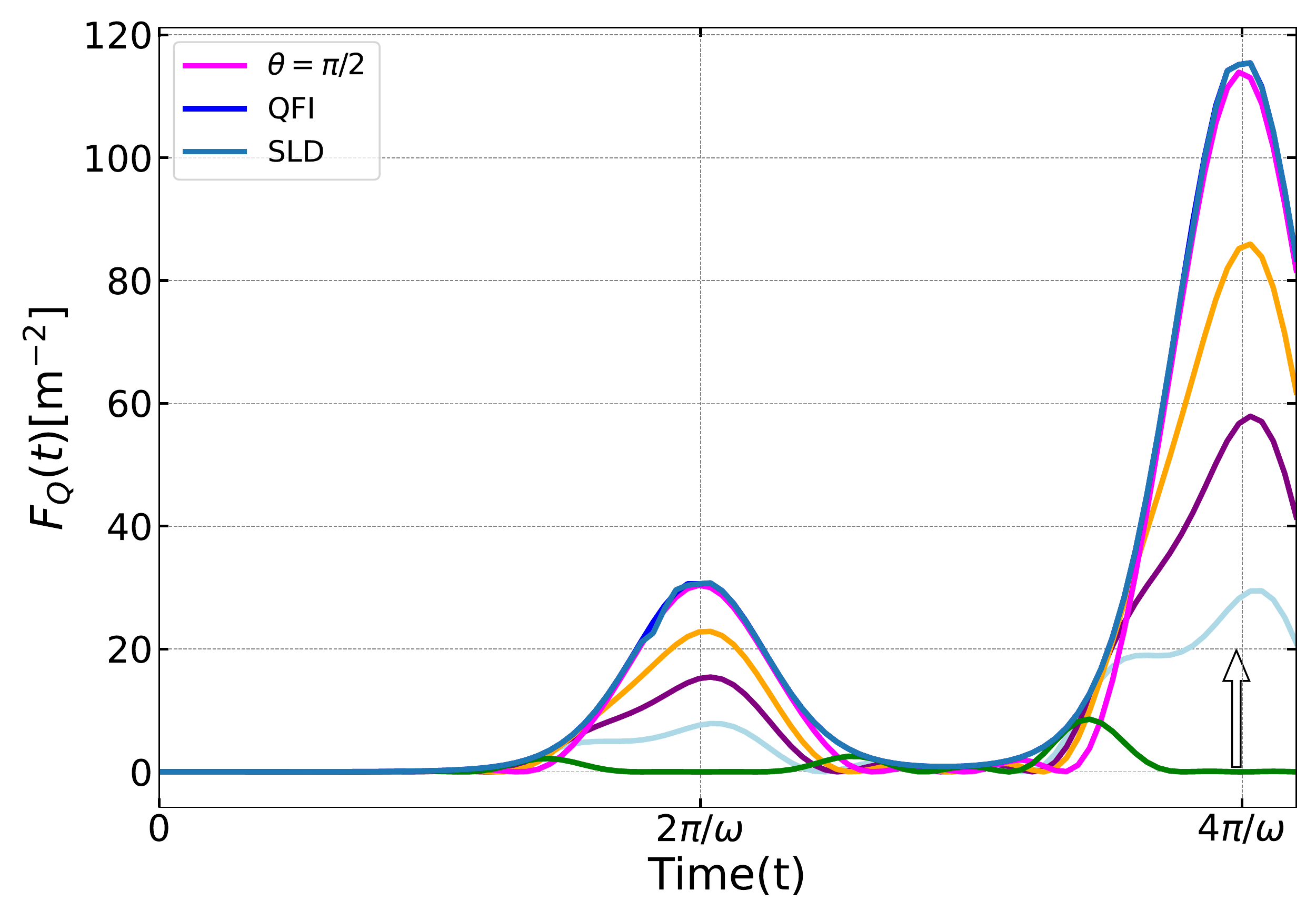}
\caption{Comparison between the QFI (Eq.~\ref{eq:qfi-sld}) and classical FI (Eq.~\ref{eq:fi}) for homodyne detection of the optical field, the values of $\theta=\{0,\pi/6,\pi/4,\pi/3,\pi/2\}$, corresponding values on the plot are indicated with an arrow. We observe that for $\theta=\pi/2$ classical FI almost reaches the QFI values if the measurement is performed at the decoupling time.}
\label{fig:fi-qfi}
\end{figure}
In the optomechanical setup, a suitable measurement candidate is homodyne detection $x = (ae^{-i\theta}+a^\dagger e^{i\theta})/\sqrt{2}$. We then numerically calculate the Classical Fisher Information (FI) to see if there are certain values of $\theta$ when classical FI reaches the QFI, where we define the classical Fisher information as: 
\begin{equation}\label{eq:fi}
    F=\int\frac{1}{p(x|d)}\Big(\frac{\partial p(x|d)}{\partial d}\Big)^2,
\end{equation}
\vspace{0.1cm}

\noindent where $p(x|d)=\text{Tr}[\rho_d\Pi_x]$ is the conditional probability with $\Pi_x$ an element of POVM. We use the eigenvectors of the operator $x$ to construct a POVM  $\Pi_x=|x\rangle\langle x|$.

We calculate the projective measurement on the eigenbasis of the SLD operator and homodyne detection measurement. Fig. \ref{fig:fi-qfi} shows that the SLD optimal measurement indeed saturates the QCRB inequality, we also find that the homodyne detection for $\theta=\pi/2$ gives a close enough result when a measurement performed at the decoupling times. 

\section{CONCLUSION}

In this work, we proposed a scheme for displacement measurement based on an opto-magno-mechanical system where a moving mirror is attached to the surface of a diamagnetically levitated graphite plate. We explored the fundamental bounds on the sensitivity $\Delta d$ of estimating the separation distance between the two magnet layers forming a magnetic trap for the graphite plate. We analyze the displacement measurement presision bounds in terms of QFI, and find that for a pure state of the cavity with $|\alpha|^2=10^6$ photons inside, the sensitivity reaches $\Delta d\sim36\times10^{-21}\text{m}$. Next, we studied the evolution of QFI of the system incorporating the decoherence in terms of photon loss. We performed numerical simulations to calculate the QFI time dependence and found that the photon leakage from the cavity causes a reduction in the QFI values at the decoupling times, which again grows with time. This implies that measurement precision is affected by the decoherence process, that reduces the pure state precision of $\Delta d\sim36\times10^{-21}\rm m$, however the presence of more photons in the cavity improves the QFI.

\begin{table*}[h!]
    \small
    \centering
    \begin{tabular}{||c |c| c||} 
        \hline
        Symbol & Parameter & Value 
        \\ [0.5ex] \hline\hline
        $\omega/2\pi$ & Mechanical frequency & $117$ Hz \\
        $\chi/2\pi$ & Optomechanical coupling rate & 55 kHz \\
        $S/2\pi$ & Gravitational coupling constant & $8\times10^{11}$ Hz \\
        $\omega_c/2\pi$ & Optical cavity frequency & $10^{14}$ Hz \\
        $d$ & Nominal vertical separation between magnet arrays & $0.25$ mm \\
        dim[$\mathcal{H}$] & Hilbert space truncation size & $50$ \\
        $(\text{l,w,d})$ & Graphite plate dimensions & $(0.1,0.1,0.04)$ mm \\
        $L$ & Optical cavity length & 100\, $\mu$m \\
        $|\alpha|^2$ & Nominal number of cavity photons & $10^6 $ \\
        $h$ & Magnets are cubes of side length & 5 mm \\
        $M_0$ & Magnetization of magnets & 1.48 Tesla \\
        $\chi_{gr}$ & Magnetic susceptibility of graphite along $(x,y,z)$ \cite{Romagnoli2022} & $-(85,85,450)\times 10^{-6}$
        \\ \hline
    \end{tabular}
    \caption{\label{tab:param} Values of physical parameters used in the work.}
\end{table*}

\begin{acknowledgments}
This work was supported by the Okinawa Institute of Science and Technology Graduate University. We are grateful for the help and support provided by the Scientific Computing and Data Analysis section of Research Support Division at OIST. We thank H. Hiyane and P. Romagnoli for useful discussions.
\end{acknowledgments}

\bibliography{main}

\end{document}